\documentclass[prd,aps,twocolumn,twoside,floatfix,showpacs]{revtex4}

\usepackage{graphicx}
\usepackage{amssymb}

\newcommand{\eg}{, {\it e.g.},\ }
\newcommand{\ie}{, {\it i.e.},\ }

\newcommand{\beq}{\begin{equation}}
\newcommand{\eeq}{\end{equation}}
\newcommand{\bea}{\begin{eqnarray}}
\newcommand{\eea}{\end{eqnarray}}

\newcommand\eqn[1]{(\ref{eq:#1})}
\newcommand\eqns[2]{(\ref{eq:#1})--(\ref{eq:#2})}
\newcommand\fig[1]{Fig.~\ref{fig:#1}}

\newcommand\sect[1]{Section \ref{sec:#1}}
\newcommand\figurewidth{0.9\linewidth}

\begin{document}

\title{Effects of mirror dark matter on neutron stars}

%%%%%%%%%%%%%%%%%%%%%%%%%%%%%%%%%%%%%%%%%%%%%%%%%%%%%%%%%%%%%%%%%%%%%%%%%%%%%%

\author{F.~Sandin}   
\email{fredrik.sandin@gmail.com}
\affiliation{IFPA, D\'epartement AGO, Universit\'e de Li\`ege, Sart Tilman, 4000 Li\`ege, Belgium}
\author{P.~Ciarcelluti}   
\email{paolo.ciarcelluti@ulg.ac.be}
\affiliation{IFPA, D\'epartement AGO, Universit\'e de Li\`ege, Sart Tilman, 4000 Li\`ege, Belgium}

%%%%%%%%%%%%%%%%%%%%%%%%%%%%%%%%%%%%%%%%%%%%%%%%%%%%%%%%%%%%%%%%%%%%%%%%%%%%%%

\begin{abstract}
If dark matter is made of mirror baryons, they are present in all
gravitationally bound structures.
Here we investigate some effects of mirror dark matter on neutron stars
and discuss possible observational consequences.
The general-relativistic hydrostatic equations are generalized to
spherical objects with multiple fluids that interact by gravity.
We use the minimal parity-symmetric extension of the standard model,
which implies that the microphysics is the same in the two sectors.
We find that the mass-radius relation is significantly modified in the
presence of a few percent mirror baryons.
This effect mimics that of other exotica\eg quark matter.
In contrast to the common view that the neutron-star equilibrium
sequence is unique, we show that it depends on the relative number of
mirror baryons to ordinary baryons.
It is therefore history dependent.
The critical mass for core collapse\ie the process by which neutron
stars are created, is modified in the presence of mirror baryons.
We calculate the modified Chandrasekhar mass and fit it with a polynomial.
A few percent mirror baryons is sufficient to lower the critical mass
for core collapse by $\sim 0.1$~M$_\odot$.
This could allow for the formation of extraordinary compact neutron
stars with low mass.
%
%We conjecture that there could be significant effects of mirror matter
%also on the dynamics of isolated neutron stars and on the lightcurve
%of some supernov{\ae}.
%
\end{abstract}

\pacs{11.30.Er, 26.60.Kp, 95.35.+d, 97.60.-s}

% 11.30.Er 	Charge conjugation, parity, time reversal, and other discrete symmetries 
% 26.60.Kp 	Equations of state of neutron-star matter 
% 95.35.+d 	Dark matter (stellar, interstellar, galactic, and cosmological)
% 97.60.-s 	Late stages of stellar evolution (including black holes)

\maketitle

%%%%%%%%%%%%%%%%%%%%%%%%%%%%%%%%%%%%%%%%%%%%%%%%%%%%%%%%%%%%%%%%%%%%%%%%%%%%%%

\section{Introduction}

Observations indicate that non-luminous matter contributes a significant
part of the mass in galaxies. This ``dark matter'' (DM) cannot be made of known
particles, but is of unknown type. In this paper we describe some potential
effects of DM on neutron stars (NS), which can be tested with future
observations of NS. These effects could be important also for the
interpretation of the observational results.
If DM exists, it should be present in all astrophysical objects,
unless there is an efficient and unexpected segregation mechanism. DM
can be present already in the formation processes and it can be accreted
subsequently from the environment. Both of these processes should naturally
occur, but the amount of DM that is trapped inside different objects should
depend on the nature of the DM particles, the type of objects and their
history. In general we expect that critical phases of stellar evolution
could be modified in the presence of relatively small amounts of DM.

The most studied hypothesis is that DM is made of
weakly interacting massive particles (WIMPs), which naturally appears
in supersymmetric extensions of the standard model (SM) of particle physics.
Should they exist, WIMPs accumulate in NS due to elastic
scattering on nucleons. If the mass of accreted WIMPs reaches a critical
value they become self-gravitating and form a dense core supported by
degeneracy pressure. The dark core collapses into a black hole if it
reaches the Chandrasekhar mass, $M \sim M_{Pl}^3/m_{DM}^2$. This limits
the possible density, cross section and mass of WIMPs
\cite{Goldman:1989nd,Bertone:2007ae}. See also \cite{Gould:1989gw}.
%The limits obtained this way are compatible with supersymmetric particles,
%but are in contrast with supermassive DM particles such as wimpzillas.
The expected high mass of WIMPs gives a low Chandrasekhar mass for the
DM core, they therefore do not affect the mass and radius of NS significantly.
The annihilation of WIMPs in the DM core produces heat and old NS
should therefore have a steady-state temperature of about $10^4$~K
\cite{Kouvaris:2007ay}. The coldest observed NS have temperatures
in the range $10^5-10^6$~K. It is observationally difficult
to detect NS with significantly lower temperatures, because the
luminosity scales as $T^4$. The effects of WIMPs on NS therefore
seems to be of little practical concern. WIMPs are, however, only one
class of DM particles and some observations suggest that the model
is oversimplified, see\eg \cite{Perivolaropoulos:2008ud} and references
therein.

Mirror matter is another viable DM candidate that emerges if one,
instead of (or in addition to) assuming a symmetry between bosons and
fermions -- supersymmetry, assumes that nature is parity symmetric.
It is a matter of fact that the weak nuclear force is not parity symmetric.
The main theoretical motivation for the mirror-matter hypothesis is
that it constitutes the simplest way to restore parity symmetry in the
physical laws of nature. When Lee and Yang proposed the non-parity
of weak interactions in 1956, they mentioned also the possibility to
restore parity by doubling the number of particles in the SM \cite{Lee:1956qn}.
Thereby the Universe is divided into two sectors with opposite handedness
that interact mainly by gravity. In the minimal parity-symmetric extension
of the SM \cite{Foot:1991bp,Pavsic:1974rq},
the group structure is $G\otimes G$, where $G$ is the gauge group of the SM.
In this model the two sectors are described by the same lagrangians, but
where ordinary particles have left-handed interactions, mirror particles
have right-handed interactions. Except for gravity, mirror matter
could interact with ordinary matter via so-called kinetic mixing of
gauge bosons,
or via unknown fields that carry both ordinary and mirror charges.
If such interactions exist they must be weak and we therefore
neglect them here.
Since photons do not interact with mirror baryons, or interact only
via the weak kinetic mixing, mirror matter constitute a natural
candidate for the DM in the Universe.
The study of the cosmological implications of mirror matter is
well-defined, because the microphysics of mirror matter is the
same as that of ordinary matter.
%There are only two {\it a priori} free parameters: the ratio of
%the relic mirror photon temperature to that of ordinary photons
%and the mirror baryon mass density.
Many consequences of mirror matter for particle physics and astrophysics
have been studied during the last decades.
The reader can refer to \cite{Okun:2006eb} for a review of the history
of mirror matter and a list of most relevant papers published before 2006.

Like their ordinary counterparts, mirror baryons can form atoms, molecules
and astrophysical objects such as planets, stars and globular clusters.
However, even though the microphysics is the same in both sectors,
the chemical content of the mirror sector should be different, because
the cosmology must be different. In particular, Big Bang nucleosynthesis
(BBN) requires that the mirror sector has lower temperature than the
ordinary sector \cite{Berezhiani:1995am, Berezhiani:2000gw}. This has
implications for the thermodynamics of the early Universe
\cite{Berezhiani:1995am,Ciarcelluti:2008vs} and for the key cosmological epochs.
Analytical results and numerical calculations of BBN \cite{Berezhiani:1995am,Ciarcelluti:2008vm}
show that the mirror sector should be helium dominated, and the abundance
of heavy elements is expected to be higher than in the ordinary sector.
Invisible stars made of mirror baryons are candidates for Massive
Astrophysical Compact Halo Objects (MACHOs), which have been observed
via microlensing events \cite{Blinnikov:1996fm,Foot:1999hm,Mohapatra:1999ih}.
Mirror stars contain more helium and less hydrogen, and therefore have
different properties than ordinary stars \cite{Berezhiani:2005vv}.
The consequences of mirror matter on primordial structure formation,
the cosmic microwave background and the large-scale-structure distribution
of matter have been studied \cite{Ignatiev:2003js,Berezhiani:2003wj,
Ciarcelluti:2003wm,Ciarcelluti:2004ij,Ciarcelluti:2004ik,Ciarcelluti:2004ip}.
These studies provide stringent bounds on the mirror sector and prove that
it is a viable candidate for DM. In addition, mirror DM is one of
the few potential explanations for the recent DAMA/LIBRA annual modulation
signal \cite{Bernabei:2008yi,Foot:2008nw,Ciarcelluti:2008qk}.

The properties of NS are intimately connected to the equation of state
(EoS) of matter at densities well beyond nuclear saturation density,
$n_0\sim 0.16$~fm$^{-3}$. NS therefore are natural laboratories for the
exploration of baryonic matter under extreme conditions, complementary
to those created in terrestrial experiments with atomic nuclei and
heavy-ion collisions.
A stiff EoS at high density is needed to explain NS with high
mass \cite{Champion06062008} and large $R_\infty$ \cite{Trumper:2003we}.
Heavy-ion collision data for kaon production and elliptic flow
provide an upper limit on the stiffness of the EoS \cite{Fuchs:2005zg,
Danielewicz:2002pu}. By combining these constraints it is possible
to test and exclude certain models of high-density EoS
\cite{Klahn:2006ir,Klahn:2006iw}. In this context it is important
to be aware of the potential effects of DM on the properties of NS.
A long-debated question in compact-star physics is whether quark matter
exist in the core of NS and whether there are unambiguous observables
associated with that. Other exotic states of matter\eg meson condensates
and hyperons could also exist in the core. For recent reviews, see
\cite{Weber:2007ch,Page:2006ud,Weber:2004kj}. For an example model of
hybrid stars (stars with a quark-matter core enclosed in a nuclear-matter
shell) that are consistent with present observational and experimental
constraints, see \cite{Klahn:2006iw}. A typical consequence of exotica
is that the maximum NS mass and the radii of NS with typical masses,
$M \sim 1.35$~M$_\odot$, becomes lower. Constraints on the mass
and mass-radius relation
of compact stars are therefore used as indicators for the presence/absence
of exotica in compact stars. As an example, see the claim in
\cite{Ozel:2006bv} and the counter-examples provided in \cite{Alford:2006vz}.
In this paper we show that if mirror matter (or, in principle, some
other form of stable self-interacting DM) accumulates in NS, the
equilibrium sequence could be  significantly modified and the effect
is similar to that of traditional exotic phases of matter. The NS
equilibrium sequence is directly related to the ground-state equation
of state and it is therefore commonly assumed to be unique. We show that
if DM accumulates in NS this is not necessarily the case. See also
\cite{Khlopov:1989fj}, where potential effects of mirror matter
on NS are discussed qualitatively.

\section{Compact star sequence}
\label{sec:sequence}

The structure of NS with a mirror-matter part is determined by
hydrostatic equations, which are similar to the well-known
Oppenheimer-Volkoff (OV) equations.
Here we repeat some essential steps in the derivation
of the OV equations, and then we generalize the
result to include mirror matter. The starting point is
the line element, $d\tau^2=g_{\mu\nu}dx^\mu dx^\nu$,
of static isotropic spacetime
\beq
	d\tau^2 = e^{2\nu(r)}dt^2 - e^{2\lambda(r)}dr^2 - r^2(d\theta^2 + \sin^2\theta d\phi^2),
\eeq
and the energy-momentum tensor of a perfect fluid
\beq
	T^{\mu\nu} = -p g^{\mu\nu} + (p+\rho)u^\mu u^\nu.
\eeq
Here, $p$ is the pressure and $\rho$ the energy
density, which includes rest-mass energy.
The Einstein field equation simplifies to
(units are chosen such that $G=c=1$)
\bea
	e^{-2\lambda(r)} &=& 1 - \frac{2M(r)}{r}, \label{eq:einstein1} \\
	\frac{d\nu}{dr} &=& \frac{\left[M(r)+4\pi r^3 p(r)\right]}{r\left[r-2M(r)\right]}, \label{eq:einstein2} \\
	\frac{dp}{dr} &=& -\left[p(r)+\rho(r)\right]\frac{d\nu}{dr}, \label{eq:einstein3}
\eea
where
\beq
	M(r) \equiv 4\pi\int_0^r d\tilde{r}\,{\tilde{r}}^2 \rho(\tilde{r}). \label{eq:mass}
\eeq
The metric should match the Schwarzschild
solution at the surface of the star, which is
located at $r=R$. It then follows from \eqn{einstein1}
that the gravitational mass of the star is $M=M(R)$.
Similarly, the differential equation \eqn{einstein2}
for $\nu(r)$ is subject to the Schwarzschild boundary
condition, so the integration constant is fixed.
In combination with an EoS,
$\rho = \rho(p)$, equation \eqn{einstein3} determines
how the pressure and density varies with $r$ inside
a star. Given an EoS and a central pressure, $p(r=0)$,
\eqn{einstein2}-\eqn{mass} are integrated until the
surface is reached, where $p(r=R)=0$. By varying the
central pressure, a one-parameter sequence of equilibrium
solutions with different M and R is obtained.

Newtonian physics applies to systems where
rest-mass energy dominates. In general relativity,
all forms of energy are sources of gravity. In
particular, the curvature of spacetime inside
a compact star depends not only on the
energy-density distribution of matter, but also
on the pressure, see \eqn{einstein2}. In the
presence of mirror matter the metric is affected
by the energy content in both sectors
\bea
	p(r) &=& p_O(r) + p_M(r), \label{eq:pressure}\\
	\rho(r) &=& \rho_O(r) + \rho_M(r). \label{eq:density}
\eea
Here, $O$ ($M$) refers to the ordinary (mirror)
sector. The metric functions, $\lambda(r)$ and
$\nu(r)$, applies to both sectors and are given
by \eqn{einstein1}-\eqn{einstein2} with the
replacements \eqn{pressure}-\eqn{density}.
Equation \eqn{einstein3} separates for the
two sectors
\bea
	\frac{dp_O}{dr} &=& -\left[p_O(r)+\rho_O(r)\right]\frac{d\nu}{dr}, \label{eq:einstein3O} \\
	\frac{dp_M}{dr} &=& -\left[p_M(r)+\rho_M(r)\right]\frac{d\nu}{dr}, \label{eq:einstein3M}
\eea
because here we assume that particles and mirror particles
interact by gravity only \footnote{
To understand this, consider the Newtonian hydrostatic equation,
$\frac{dp}{dr} = -\rho_0(r)\frac{d\Phi}{dr}$,
which tells that the gradient of the pressure
is equal to the rest-mass density, $\rho_0$,
of fluid elements times the gradient of the
gravitational potential, $\Phi$. This is a
classical force-balance equation. The gradient
of the pressure in the mirror sector does not
exert a direct force on fluid elements in the
ordinary sector, and vice-versa, because here
we assume that matter and mirror matter interact
by gravity only. In the general-relativistic equation
\eqn{einstein3} the inertia of fluid elements is
$p(r)+\rho(r)$ and $\nu(r)$ is the generalization
of the gravitational potential. The separation
of \eqn{einstein3}, with the replacements
\eqn{pressure}-\eqn{density}, into
\eqn{einstein3O}-\eqn{einstein3M} follows because
the two sectors interact by gravity only.}.
This result can be generalized to
any number of fluids that interact by gravity. 
In principle the two sectors could interact by other forces
than gravity\eg by photon--mirror-photon kinetic coupling,
which gives charged mirror particles an effective electric
charge in the ordinary sector that is about eight or nine orders
of magnitude lower than that of ordinary charged particles
\cite{Foot:2000vy}.
The effect of such eventual weak interactions
on the equilibrium structure of a compact star should be
small and we therefore consider only the gravitational
interaction here. 

\begin{figure}
	\includegraphics[width=\figurewidth,clip=]{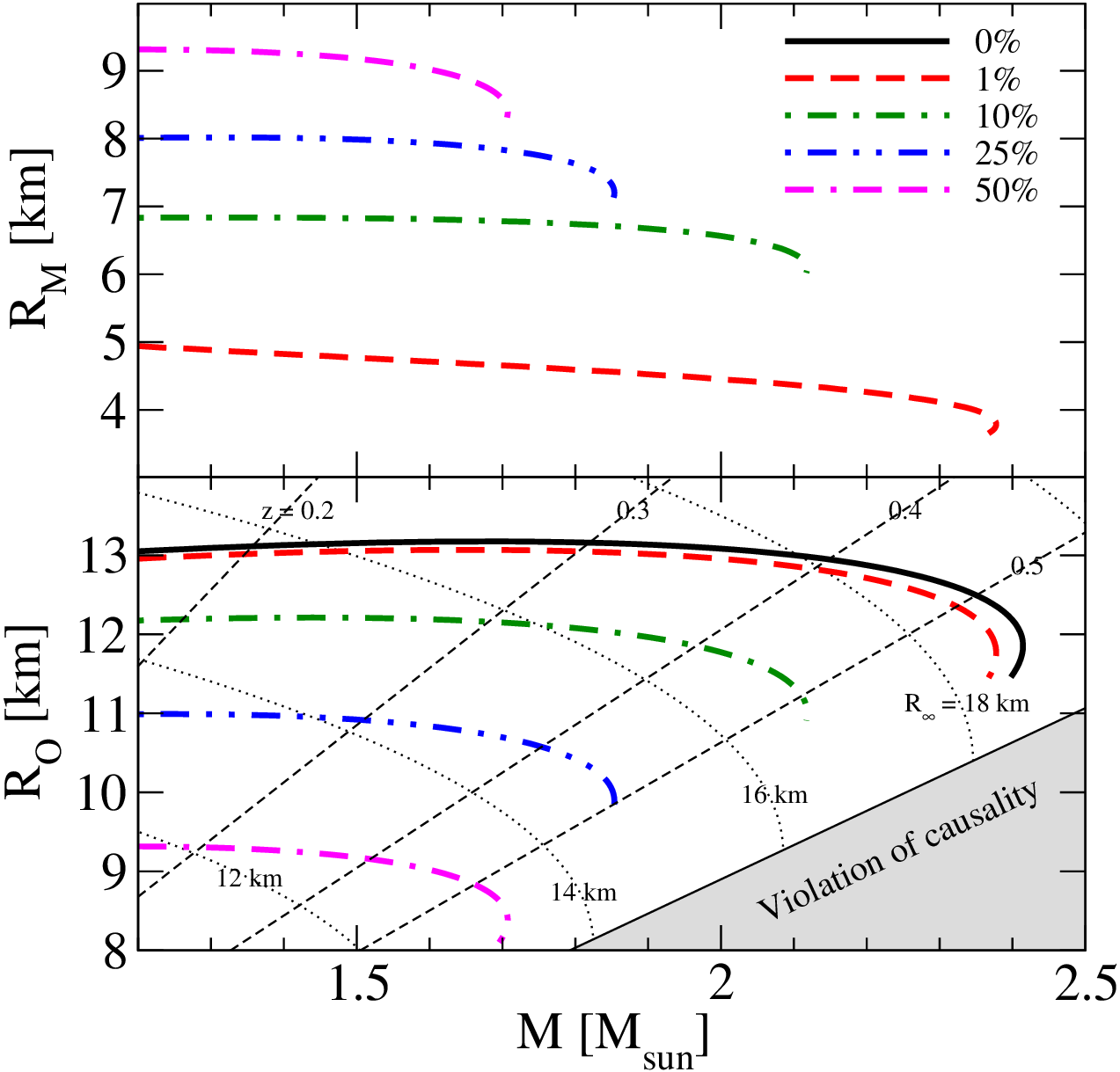}
	\caption{(Color online) Neutron-star sequences for
	different number of baryons in the mirror sector, $N_M/(N_O+N_M)=$~0\%,
	1\%, 10\%, 25\% and 50\%. Here, $M$ is the total
	gravitational mass of the star and $R_O$ ($R_M$) is the
	coordinate radius of the surface in the ordinary (mirror)
	sector. Here we consider $N_M\leq N_O$ only, but the results
	can be generalized to the opposite case by the replacement
	$R_O \leftrightarrow R_M$.
	%\ie a $1.4$~M$_\odot$ mirror neutron star with 1\% ordinary
	%baryons would have an observable radius of about 5~km.
	These sequences correspond to a nuclear relativistic
	mean-field equation of state \cite{Typel:2005ba} and the
	crust is modeled with the BPS equation of state \cite{Baym:1971pw}.
	Included in the figure are also lines of constant surface
	redshift (thin dashed lines), $z = 1/\sqrt{1-2M/R}-1$, and
	``radiation radius'' (thin dotted lines), $R_\infty = R/\sqrt{1-2M/R}$.
	The latter quantity can be constrained via the surface emission
	from isolated neutron stars and is, in addition to the mass, an
	important indirect observable \cite{Trumper:2003we}. The surface
	redshift has not yet been reliably observed, but with future
	high-resolution observatories and a better understanding of
	neutron-star atmospheres this can become an important observable.
	\label{fig:seq}
	}
\end{figure}

Equations \eqn{einstein2} and \eqns{mass}{einstein3M}
are the general-relativistic hydrostatic equations for
spherical stars with a mirror matter part.
By choosing different central pressures in
the two sectors, a star with different radius and
baryon number in the two sectors is obtained.
In \fig{seq} we plot compact star sequences with different
number of baryons in the mirror and ordinary sectors.
The EoS used here is based on a relativistic mean-field
model of nuclear matter \cite{Typel:2005ba}, which
is combined with the Baym-Pethick-Sutherland (BPS)
EoS for the crust \cite{Baym:1971pw}.
In this work we approximate the EoS of mirror nuclear
matter with that of ordinary nuclear matter\ie
we use the same EoS in the two sectors.
We motivate this approximation in the following way:
NS are formed from the iron core of stars\ie of nuclei with
maximum binding energy formed at the end of nuclear burning.
They quickly become cold on the nuclear energy scale,
because of neutrino emission, and matter thereafter
is in the ground state. The situation is different for the
mirror-matter part of an NS. Different types of mirror
nuclei can be accreted\eg from a binary companion mirror
star, and an eventual mirror core of the progenitor star
is not necessarily made of iron.
The mirror core is, however, compressed into a compact
object by the strong gravitational field of the NS.
Mirror nuclei are therefore disintegrated into mirror
neutrons, mirror protons and light mirror nuclear
clusters, which should equilibrate with respect to weak
nuclear reactions, just like neutrons and protons do
in ordinary NS.
In the minimal parity-symmetric extension of the SM
(see the Introduction) the lagrangian of the mirror
sector is identical to that of the ordinary sector.
In this context it therefore seems reasonable to approximate
the EoS of high-density mirror nuclear matter with that
of ordinary nuclear matter.

\begin{figure}
	\includegraphics[width=\figurewidth,clip=]{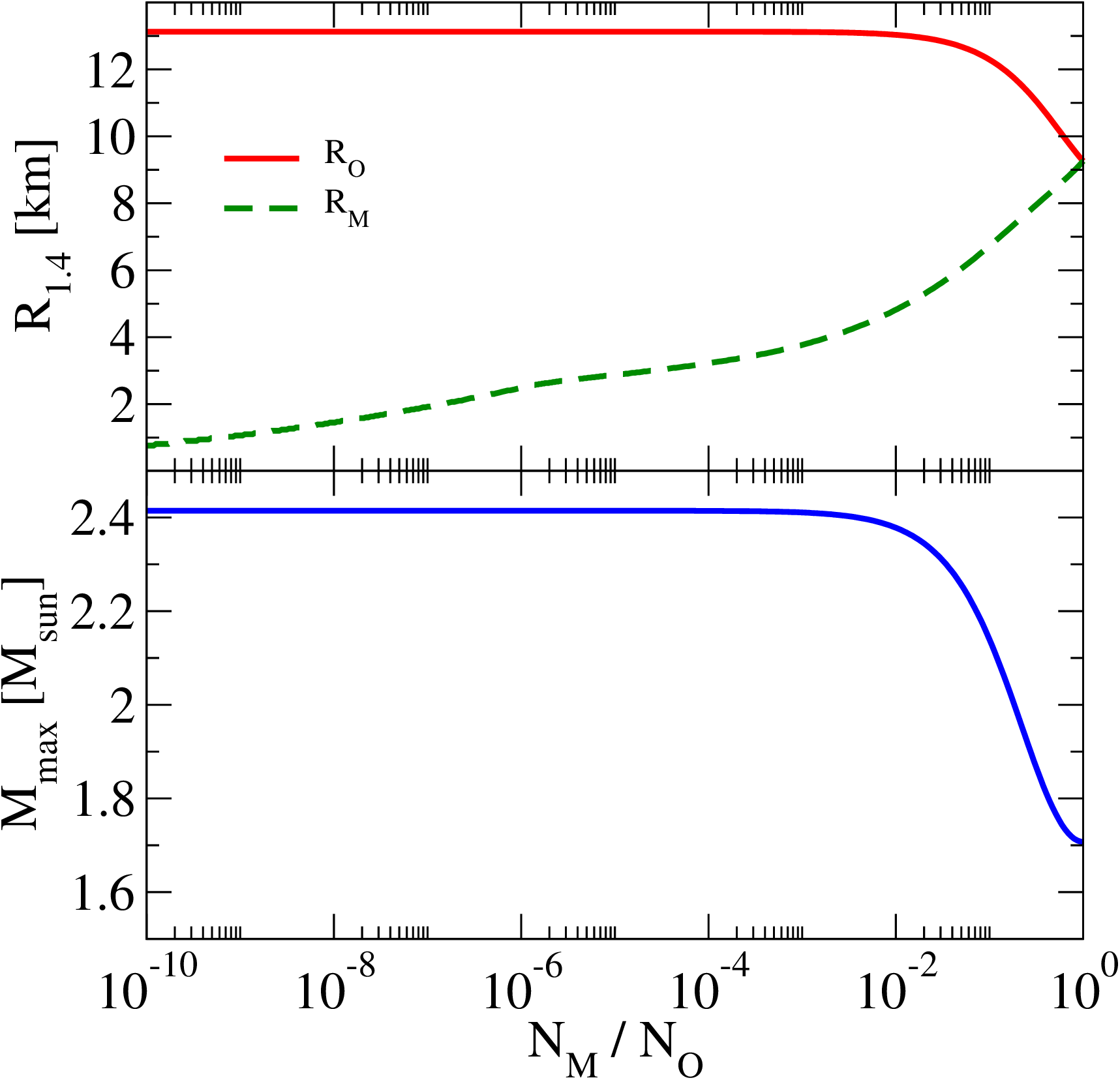}
	\caption{(Color online) Radii of $1.4$~M$_\odot$ stars (upper
	panel) and maximum masses (lower panel) vs. the relative
	number of mirror baryons to ordinary baryons.
	%These results correspond to a nuclear relativistic
	%mean-field equation of state \cite{Typel:2005ba}.
	\label{fig:maxmass}}
\end{figure}

\begin{figure}
	\includegraphics[width=\figurewidth,clip=]{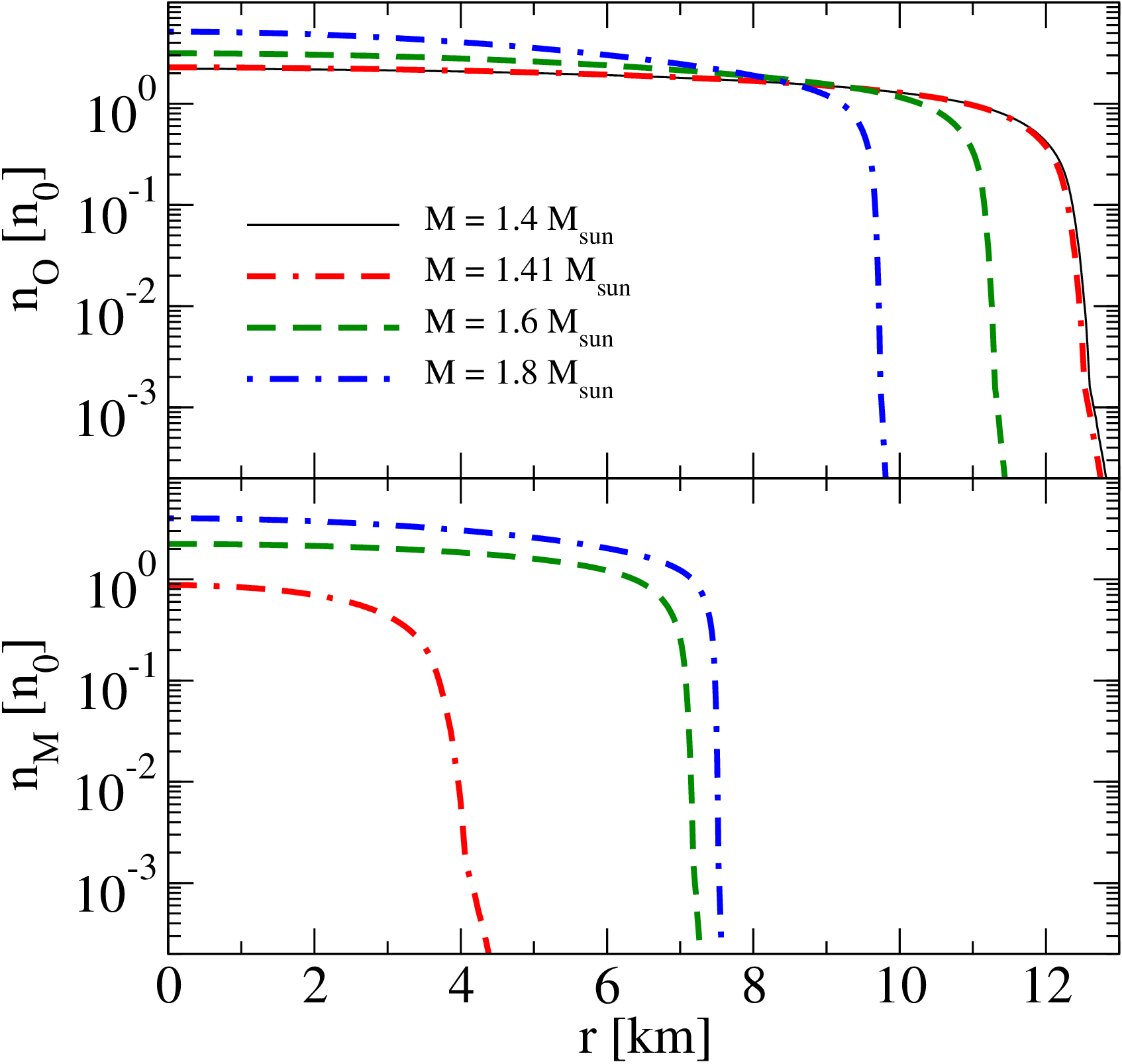}
	\caption{(Color online) Density profiles of neutron stars
	with identical baryon number in the ordinary sector,
	but different number of mirror baryons. The number of
	ordinary baryons is fixed such that $M=1.4$~M$_\odot$
	in absence of mirror matter. \label{fig:profiles}}
\end{figure}

The number of baryons in a star is the integral over
the invariant volume element, $\sqrt{\text{det}(g_{\mu\nu})}d^4x$,
and the conserved baryon number current. The standard
textbook result is
\bea
	N_{O,M} = 4\pi\int_0^R dr\,r^2 n_{O,M}(r)
	\left[1-\frac{2M(r)}{r}\right]^{-1/2}, \label{eq:number}
\eea
where $n_O$ ($n_M$) is the number density of
baryons (mirror baryons). The Schwarzschild
boundary condition applies to the surface at
$\text{max}(R_O,\,R_M)$, because the local
curvature of empty space in one sector is
affected by the matter content in the other
sector. The coordinate radius and the maximum
mass of the sequence depend on the relative
baryon number in the two sectors, see \fig{maxmass}.
In \fig{profiles} we illustrate the density profiles
of NS with identical number of ordinary baryons and
different number of mirror baryons.

From these results it is clear that the mass-radius
relation of NS is significantly modified in the presence
of a few percent mirror baryons.
The results presented in \fig{maxmass}
can be generalized to $N_M>N_O$ by the replacement
$M \leftrightarrow O$. It then follows that a
mirror NS could be accompanied by a
compact object in the ordinary sector that is only
a few kilometers in size. Should they exist, such
objects would have extraordinary properties, as
the net gravitational mass is comparable to
that of an ordinary NS. The relevance
of these results in the mirror matter scenario
depends on the probability for significant amounts
of mirror baryons (baryons) to accumulate in the gravitational
potential of (mirror) NS and their progenitor stars.

\section{Core collapse and limiting mass}

The critical mass of the core of ordinary stars is modified
in the presence of mirror matter, because mirror particles
act as an additional source of gravity. Such stars therefore
collapse at lower core masses and this could result in
extraordinary compact NS with low mass, due to the
presence of mirror matter. In order to estimate the magnitude
of the effect we modify the standard textbook calculation of
the Chandrasekhar mass of a gravitating object supported by
electron degeneracy pressure, see\eg \cite{Camenzind:2007}
or \cite{Glendenning:2000cs}. We are interested in cores with
high density that are near gravitational collapse. The density
is essentially given by the rest-mass density of nucleons,
$\rho_0$, and the dominant contribution to the pressure, $p$,
comes from ultrarelativistic electrons. The EoS is therefore
approximately polytropic, $p = K \rho_0^\gamma$, with $\gamma=\frac{4}{3}$.
The factor $K$ depends on the relative number of electrons
to nucleons and therefore depends on the chemical composition
of the core. To leading order, this EoS and the hydrostatic
equations give a limiting mass of
\beq
	M_c \simeq 1.46 M_\odot \left(\frac{Y_e}{0.5}\right)^2,
\eeq
where $Y_e=n_e/n_B$ is the electron fraction\ie the relative
number density of electrons to baryons. This is the
traditional expression for the Chandrasekhar mass.

\begin{figure}
	\includegraphics[width=\figurewidth,clip=]{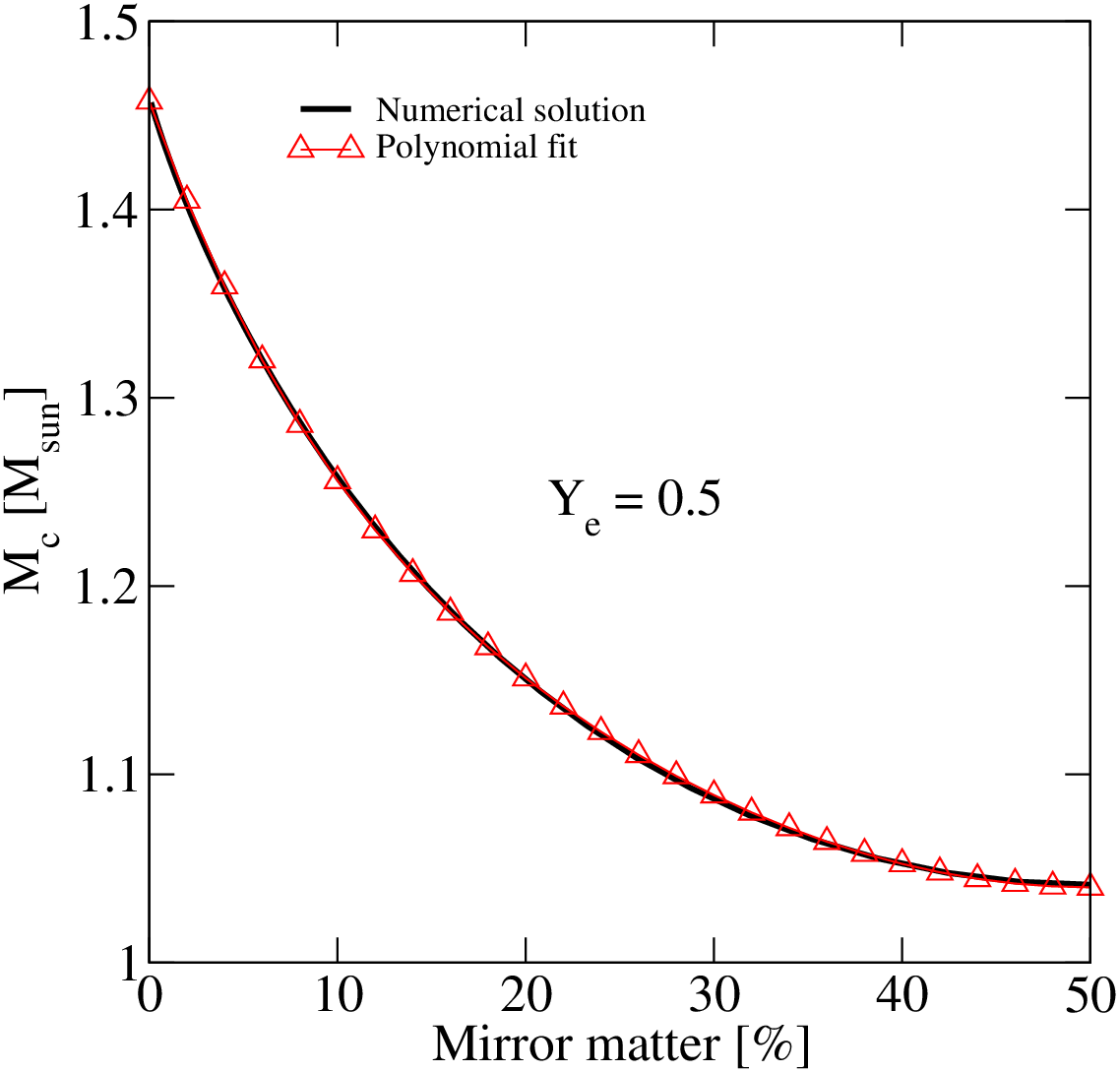}
	\caption{(Color online) %Critical mass of the core of an ordinary star
	The modified Chandrasekhar mass vs. the relative number of baryons
	in the mirror sector, $N_M/(N_O+N_M)$, at fixed electron fraction, $Y_e=0.5$.
	In the limit of 0\% mirror baryons the traditional Chandrasekhar
	mass, $M_c \simeq 1.46$~M$_\odot$ is reproduced. The critical
	mass depends on the electron fraction\ie the chemical composition,
	which could be different in the two sectors. For equal electron
	fractions the $Y_e$-dependence of the model is simple, see \eqn{mcfit}.
	\label{fig:masslimit}}
\end{figure}

A self-consistent derivation of the limiting mass requires that
the hydrostatic equations in the two sectors are coupled and
solved simultaneously, as in \sect{sequence}. This is not
trivial because the radial dependencies of the pressure and
density are different in the two sectors.
In order to obtain the critical mass as a function of the
relative number of baryons in the two sectors we solve
the hydrostatic equations, \eqn{einstein2} and \eqns{mass}{einstein3M},
with the polytropic EoS used in the derivation of the
ordinary Chandrasekhar mass. The result is shown in \fig{masslimit}.
It is not necessary to account for general-relativistic effects in
calculations of the structure of ordinary (and white-dwarf) stars,
but since the equations are already presented in \sect{sequence}
we use them.

We limit the discussion here to the simplified situation when the
electron fractions in the two sectors are equal. In this case
the leading-order $Y_e$-dependence in the standard expression for
the Chandrasekhar mass applies also to the situation when mirror
baryons are present in the core\ie the numerical solution shown
in \fig{masslimit} is proportional to $(Y_e/0.5)^2$.
By fitting a polynomial to the numerical results we obtain an
expression for the modified Chandrasekhar mass
\beq
	M_c \simeq
	\left(1.04 \!+\! 1.26 q^2 \!-\! 1.36 q^4 \!+\! 12.0 q^6\right)\!
	\left(\frac{Y_e}{0.5}\right)^2\!\!\text{M}_\odot,
	\label{eq:mcfit}
\eeq
where $q=N_M/(N_O+N_M)-0.5$ is the relative number of mirror
baryons shifted by an offset of $-50$\% that necessarily
makes the solution even, because we use the same EoS in both
sectors. The fit is shown in \fig{masslimit}. In the limit of
0\% mirror baryons the traditional Chandrasekhar mass,
$M_c \simeq 1.46$~M$_\odot$ is reproduced. For 50\% mirror
matter our result is consistent with the estimate in
\cite{Kolb:1985bf}. The nature of
the critical mass is somewhat different than the picture in
this simplified model. In particular, the real cause of the
collapse is the capture of energetic electrons by protons and
a softening of the EoS in the core.
The relativistic mass-limit considered here gives a qualitative
picture of how the critical mass of the core is modified when
mirror DM is present, or is accreted from the environment.

\section{Accretion of mirror matter}
\label{sec:accretion}

The distribution of mirror DM in galaxies is expected to
be non-homogeneous, because mirror baryons should form complex
structures in a similar way as ordinary baryons do.
There are essentially three different possibilities for the
capture of mirror matter by an ordinary NS (the same would
apply for the opposite situation\ie capture of ordinary matter
by a mirror NS). First, NS should accrete particles from the mirror
interstellar medium and this must not be in conflict with
observations\eg by leading to gravitational collapse of young
NS or too high surface temperatures vs. estimated ages.
Secondly, the accretion rate could be significantly enhanced if
an NS passes through a high-density region of space\eg a
mirror molecular cloud or planetary nebula. The third possibility
is that NS merge with macroscopic bodies in the mirror sector,
causing violent events and possibly a collapse into a black hole.
The probability of such events is low and should decrease with
the mass of the dark objects, but they could be relatively
easy to discover due to the high energy output\eg in the form
of strange supernova-like events. One such observation has
recently been reported \cite{Barbary:2008ge}. We expect that
the accretion rate increases towards the center of the galaxy and
where the concentration of dark matter is high\eg in mirror
stellar clusters.
Therefore, the effects of mirror matter should depend on the
location and history of each star. Note that in
general the local environment in the hidden sector is different
from that in the visible sector. For example, an NS located
in an empty galactic region could be surrounded by stars or a
molecular cloud in the mirror sector. In fact, the structure
formation process is different in the two sectors. Structures
at small scales, like stars and stellar clusters, are formed
essentially independently in the two sectors because they are
electromagnetically decoupled. The process of accretion of
mirror matter in ordinary stars were first studied by Blinnikov
and Khlopov many years ago \cite{Blinnikov:1983itep126,Blinnikov:1983gh}.

The accretion rate of mirror matter depends on the location
of the NS and the structure of the hidden mirror sector, which
is unknown.
We therefore limit the discussion here to some rough upper-bound
estimates, which nevertheless leads to a useful conclusion.
The accretion rate of mirror baryons can be estimated with the
result for capture of collisionless particles in the gravitational
potential of a compact object, Eq.~(14.2.21) in \cite{Shapiro:1983du}:
\beq
	\dot{M} = \left(\frac{\rho_\infty}{10^{-24}\,\text{g}/\text{cm}^{3}}\right)
	\left(\frac{10\,\text{km}/\text{s}}{v_\infty}\right)
	\left(\frac{M}{\text{M}_\odot}\right)^2
	\text{kg}/\text{s}.
	\label{eq:accretionrate}
\eeq
Here $\rho_\infty$ and $v_\infty$ are, respectively, the density
and speed of mirror particles. In the derivation of this expression
it is assumed that the distribution of mirror particles is isotropic
and monoenergetic at large distance from the NS. The quantities
are normalized to the typical values for the interstellar medium
in the ordinary sector of our galaxy.
% The particle speed is roughly related to the temperature by
% $\frac{3}{2}k T=\frac{1}{2}m v_\infty^2$.
Giant molecular clouds made of ordinary baryons can be tens of parsecs
in size and the particle number density can reach $10^6$~cm$^{-3}$ in
some regions \cite{Ferriere:2001rg}.
The clouds consist mainly of hydrogen ($\sim 70$\%) so the density
can reach $10^{-18}$~g/cm$^3$. If the density of the interstellar
medium in the mirror sector reaches this value, an NS at that location
would accrete about $10^6$~kg/s. It is possible that mirror molecular
clouds have higher density than their ordinary counterparts, because
the mirror sector is subject to different initial conditions and should
have different chemical composition. In particular, it is expected to be
helium dominated. We account for these uncertainties by increasing the
upper-limit estimate for the accretion rate from the mirror
interstellar medium to $10^7$~kg/s. We have checked that this limit
remains valid also for NS with arbitrary high kick-velocities.
In fact, the accretion rate is lower for NS with high speed.
An accretion rate of $10^7$~kg/s is, however, far too low to have
a significant effect on the mass and structure of NS. Mirror matter must
therefore originate from the progenitor star, or from a companion mirror star,
if the effects on the mass-radius relation discussed in
\sect{sequence} are to be observable. Note that
the density considered above exceeds the average dark-halo density,
which varies between about a tenth of a GeV/cm$^3$ to hundreds of
GeV/cm$^3$ depending on the model used and the location in the galaxy.
One GeV/cm$^3$ corresponds to about $10^{-24}$~g/cm$^3$, so on
average the dark halo density is well below the $10^{-18}$~g/cm$^3$
that we consider in the estimate above.

\begin{figure}
	\includegraphics[width=\figurewidth,clip=]{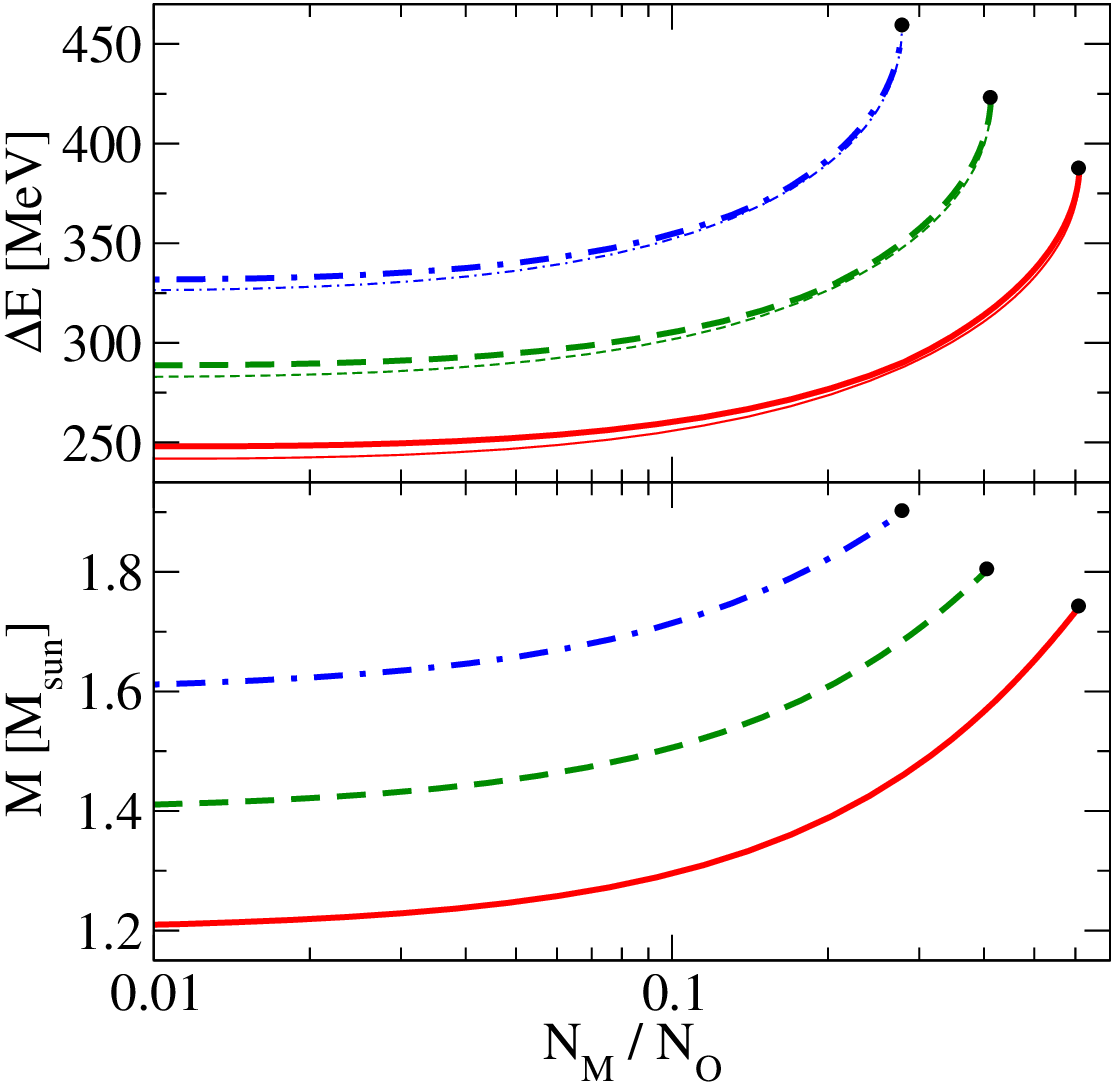}
	\caption{(Color online) Change of binding energy per accreted
	mirror baryon (upper panel) and the total gravitational mass
	(lower panel) vs. the relative number of mirror baryons.
	The baryon number in the ordinary sector is fixed such that
	$M(N_M=0)=1.2$, $1.4$ and $1.6$~M$_\odot$. Bold lines indicate
	the total change of binding energy \eqn{binding}, which
	includes both gravitational binding energy and nuclear
	binding energy. Thin lines indicate the kinetic energy
	of a mirror baryon that falls from great distance to
	the surface of the mirror-matter core, see text for
	further information. The solid disks indicate maximum-mass
	configurations\ie stars with higher mirror-baryon number
	are gravitationally unstable.
	\label{fig:binding}}
\end{figure}

The gravitational binding energy of NS is of the order of
10\% of the mass-energy. While the accretion rate is too
low to significantly affect the mass and structure of NS,
one could imagine that the associated release of gravitational
binding energy heats the star. The total binding energy can
be calculated by comparing the energy of an NS in equilibrium
before and after the accretion of a mirror baryon.
The total mass-energy of a star is $M$ (in units where $c=1$)
and the binding energy is $m_B N_B - M$, where $M$ is
the gravitational mass and $m_B$ ($N_B$) is the baryon mass
(net baryon number). The change of the binding energy when
the mirror baryon number is increased by one is
\beq
	\frac{\partial E_B}{\partial N_M}
	= \left.\frac{\partial}{\partial N_M}\right|_{N_O}\!\!\!\!\!\!\left(m_B N_B - M\right)
	= m_B - \left.\frac{\partial M}{\partial N_M}\right|_{N_O}.
	\label{eq:binding}
\eeq
This quantity is plotted with bold lines in the upper panel
of \fig{binding} for three different choices of the baryon
number in the ordinary sector. The EoS used is the same as
that in \sect{sequence}.
Most of this binding energy is released in the mirror sector.
The kinetic energy of a mirror baryon that falls from great
distance to the mirror-matter surface is $m_B[1-\sqrt{g_{00}(R_M)}]$,
where $g_{00}(R_M)$ is the time-component of the metric tensor
at the surface.
This kinetic energy is plotted with thin lines in the upper
panel of \fig{binding} and it accounts for more than 97\% of
the total binding energy in all three cases.
A significant part of the remaining small binding energy is
accounted for by in-medium binding of the mirror baryon\ie
$m_B$ is higher than the energy per baryon in the medium.
In principle there can be some heating of the ordinary part of
the NS\eg due to weak interactions between the two sectors,
but as we demonstrate below such effects are irrelevant from
an observational point of view. See also \cite{Blinnikov:2009nn},
where this is discussed in some detail.

Isolated NS have essentially black-body spectra, so the net
surface emission is given by the Stefan-Boltzmann law,
$\dot{E} = 4\pi\sigma T^4 R^2$. The observed surface temperature
of NS is around $10^6$~K and the radius is of the order $10$~km,
so the radiated energy is about $10^{25}$~J/s \footnote{Note
that young neutron stars cool mainly by neutrino emission. The
transition from neutrino-dominated cooling to photon-dominated
cooling occurs at an age of about 100,000 years.}.
If we assume that all binding energy is released as heat
in the ordinary sector (this is clearly an overestimate),
the upper-limit for the accretion rate, $10^7$~kg/s,
would correspond to a total heating effect of about $10^{23}$~J/s.
The thermal emission from the coldest observed NS is about
two orders of magnitude higher than this upper-limit estimate
for the heating.
Furthermore, most of the binding energy is converted to kinetic
energy of mirror particles when they fall in the gravitational
field. These particles will heat the mirror-matter core on impact,
but that does not affect the temperature of ordinary matter,
because the two sectors are electromagnetically decoupled.
Consequently, the accretion of mirror matter
in NS does not cause significant heating of the ordinary part
of the star. The accretion-driven heating should be somewhat
higher in the case of WIMPs, but the effect is presently not
observable \cite{Kouvaris:2007ay}.

\section{Conclusions}

If DM is made of mirror baryons, they are present in all
gravitational structures. In particular, they should exist in
stars at the final stages of stellar evolution. In this paper
we show that this could have significant effects on the
formation and structure of NS. Our knowledge of NS is limited
and highly dependent on observations. The situation is
complicated by the fact that the conditions inside NS cannot
be reproduced in laboratories. Ideally, we should learn how to
calculate the ground-state EoS from first principles using the
SM and then test it with observations of NS, but this task is
presently too complicated. An intermediate goal is to unify
``low-density'' results from nuclear-physics experiments and
lattice-QCD calculations with observed properties of NS. In
that context it is important to understand the potential effects
of DM on NS, so that the observations are not misunderstood
and false conclusions about the microphysics are made. After
all, DM apparently is dominant over ordinary matter and we
do not know what it is. In this paper we illustrate that
the NS sequence is significantly modified in the presence
of a few percent mirror DM and the effect mimics that of
other exotic forms of matter\eg quark matter. An observation
of an extraordinary compact NS is therefore not sufficient
to conclude that there must be some form of exotic phase
of ordinary matter in the star, since it can be explained
also by the presence of mirror matter.

We find that the NS equilibrium sequence is a function of
the mirror DM content\ie in contrast to the equilibrium
sequence of ordinary NS it is not a one-parameter sequence.
The maximum mass of the sequence and the radii of
stars with typically observed masses, $M \sim 1.35$~M$_\odot$,
decrease with increasing mirror baryon number.
The {\it non-uniqueness} of the equilibrium sequence is
the main result of this work, because it is not clear that
it can be explained with traditional physics, nor with
self-annihilating DM candidates such as supersymmetric
particles. New measurements of NS masses and mass-radius
relations are frequently made, thanks to the high resolution
and sensitivity of modern observatories. The uniqueness
property of the NS equilibrium sequence could therefore
be put to the test in the near future.
If it turns out that NS masses and radii cannot be explained
with a one-parameter sequence (after effects of rotation are
accounted for), one could search for a correlation between
the orbits of the observed NS and the expected density of
DM along those orbits.
The accumulation of mirror matter in stars could have a
significant effect also on the critical mass for core collapse.
This is a potential formation mechanism for NS with low masses
and extraordinary small size.
These results can be used in future studies to test
models of the mirror Universe.
\newline
\vspace{0.2cm}

%It would be interesting to analyze the oscillations
%that could be induced in NS with a mirror matter part.
%Collisions in either sector would induce oscillations that
%could be observable, since the two objects are bound by gravity
%but are otherwise free to move independently of each other.

%%%%%%%%%%%%%%%%%%%%%%%%%%%%%%%%%%%%%%%%%%%%%%%%%%%%%%%%%%%%%%%%%%%%%%%%%%%%%%

\acknowledgments
%\section*{Acknowledgments}

We acknowledge support from the Belgian fund for scientific research (FNRS).
The equation of state used in this work was provided by S. Typel and a second
equation of state that we used for comparison was provided by T. Kl{\"a}hn.
We are grateful to S.~Blinnikov for useful discussions that improved
the manuscript. We thank M.~Y.~Khlopov for help with the bibliography.

%%%%%%%%%%%%%%%%%%%%%%%%%%%%%%%%%%%%%%%%%%%%%%%%%%%%%%%%%%%%%%%%%%%%%%%%%%%%%%

\bibliographystyle{h-physrev.bst}
\bibliography{ref}

\end{document}